\documentclass[prl,twocolumn,floatfix,bibnotes]{revtex4-1}

\usepackage{physics}
\usepackage{gensymb}
\usepackage{graphicx}
\usepackage{amsmath}
\usepackage[hidelinks]{hyperref}
\usepackage{newfloat}

\newcommand*\circled[1]{\raisebox{.5pt}{\textcircled{\raisebox{-.9pt} {#1}}}}

\begin{document}

\title{Interaction control and bright solitons \\ in coherently-coupled Bose-Einstein condensates }
\author{J. Sanz}
\author{A. Fr\"{o}lian}
\author{C. S. Chisholm}
\author{C. R. Cabrera}
\author{L. Tarruell}\email{Electronic address: leticia.tarruell@icfo.eu}
\affiliation
{ICFO -- Institut de Ciencies Fotoniques, The Barcelona Institute of Science and Technology, 08860 Castelldefels (Barcelona), Spain}

\begin{abstract}
We demonstrate fast control of the interatomic interactions in a Bose-Einstein condensate by coherently coupling two atomic states with intra- and inter-state scattering lengths of opposite signs. We measure the elastic and inelastic scattering properties of the system and find good agreement with a theoretical model describing the interactions between dressed states. In the attractive regime, we observe the formation of bright solitons formed by dressed-state atoms. Finally, we study the response of the system to an interaction quench from repulsive to attractive values, and observe how the resulting modulational instability develops into a bright soliton train.
\end{abstract}

\date{\today}
\maketitle
Attractive non-linear systems host rich physics. Prime examples are solitons -- self-bound states that propagate without dispersion in one-dimensional geometries -- and modulational instability -- the breakup of a periodic wave into a train of such solitons \cite{Dauxois2006, Kamchatnov2000}. Bose-Einstein condensates (BECs) with attractive interactions are an ideal platform for the controlled experimental study of these phenomena. Matter-wave bright solitons have been observed in optical waveguides using $^7$Li \cite{Khaykovich2002, Strecker2002, Medley2014}, $^{85}$Rb \cite{Cornish2006, Marchant2013, McDonald2014}, $^{39}$K \cite{Lepoutre2016}, and $^{133}$Cs atoms \cite{Meznarsic2019, DiCarli2019}. Furthermore, BECs have been exploited to study bright soliton collisions \cite{Nguyen2014}, to form quantum droplets -- higher dimensional analogues of bright solitons that are stabilized by quantum fluctuations \cite{Ferrier2016a, Chomaz2016, Cabrera2018, Cheiney2018, Semeghini2018, Ferioli2019} -- and to explore the modulational instability leading to the formation of soliton trains \cite{Nguyen2017, Everitt2017}. A key ingredient of these experiments is the ability to control the strength and sign of the interatomic interactions in a time-dependent manner. Although normally achieved using magnetic Feshbach resonances \cite{Chin2010}, the limited temporal bandwidth of this method has stimulated the development of complementary approaches based on optical fields \cite{Theis2004, Enomoto2008, Bauer2009, Clark2015, Jagannathan2016}.

In this Letter, we demonstrate an alternative method for controlling interactions in a fast and flexible manner. In our scheme, two internal states of a BEC with different scattering lengths are coherently-coupled exploiting a radio-frequency (rf) field, which modifies the scattering properties of the corresponding dressed states. Until now, this effect could only be observed indirectly through the change of miscibility in binary BEC mixtures \cite{Nicklas2011, Nicklas2015, Shibata2019}. Here we show that exploiting a system with inter- and intra-state interactions of opposite signs enables large modifications of the elastic and inelastic scattering properties of these dressed states. They can be flexibly controlled by adjusting the parameters of the coupling field, also extending to the attractive regime. In this case, we demonstrate the stabilization of bright solitons formed by dressed-state atoms. Furthermore, we exploit the high temporal bandwidth of this technique to quench the interactions from repulsive to attractive values, and observe how the resulting modulational instability develops into a bright soliton train.

\begin{figure}[t]
\centering
\includegraphics[clip,scale=0.85]{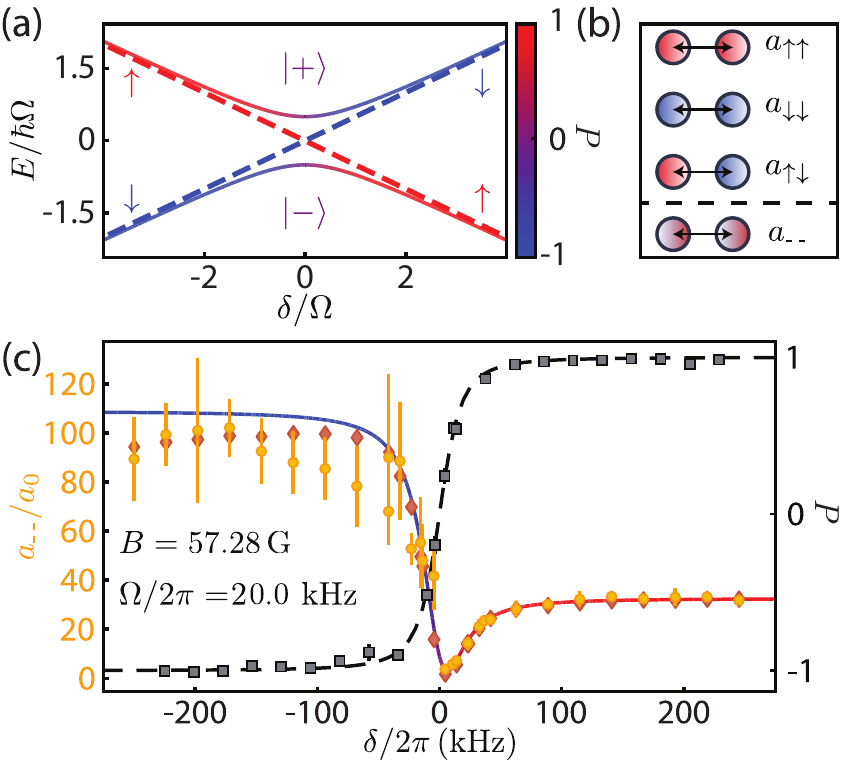}
\caption{Elastic scattering properties of the lower dressed state. (a) Energy $E$ of dressed states $\ket{-}$ and $\ket{+}$ vs. detuning $\delta$, normalized to the Rabi frequency $\Omega$. Here $\uparrow$ and $\downarrow$ are the bare atomic states. Colorscale: state composition in terms of $P=\delta/\sqrt{\Omega^2+\delta^2}$. (b) Sketch of the bare scattering lengths involved and resulting effective scattering length $a_{\text{-\,-}}$. (c) Experimental value of $a_{\text{-\,-}}$ obtained by scaling $\sigma_x^5/N$ (orange circles, left axis) and $P$ (gray squares, right axis) vs. $\delta$. Lines: theory predictions. Brown diamonds: numerical simulation of the expansion. Colorscale of the $a_{\text{-\,-}}$ curve: value of $P$. Error bars: standard deviation from $5$ independent measurements.}\label{fig1}
\end{figure}

We perform all experiments in the strong coupling limit, where the Rabi frequency of the rf field $\Omega$ dominates over all other energy scales of the problem and the system is conveniently described by the dressed states $\ket{-}=\sin{\theta}\ket{\downarrow}-\cos{\theta} \ket{\uparrow}$ and $\ket{+}=\cos{\theta}\ket{\downarrow}+\sin{\theta}\ket{\uparrow}$. We refer to them as lower and higher dressed state, respectively. The mixing angle $\theta$ fixes the composition of the system in terms of the bare states $\ket{\uparrow}$ and $\ket{\downarrow}$. It is given by $\cos^2{\theta}=(1+P)/2$, where $P=\delta/\tilde{\Omega}$ is the polarization parameter, $\delta$ the detuning of the coupling field and $\hbar \tilde{\Omega}=\hbar\sqrt{\Omega^2+\delta^2}$ is the energy gap between the two dressed states, see Fig. \ref{fig1}(a). Here $\hbar$ is the reduced Planck constant.

To describe the interactions between dressed states, we rewrite the interaction part of the Hamiltonian in the dressed-state basis \cite{Search2001}. The resulting collisional couplings account for processes which either preserve the two-particle dressed state of the colliding atoms (elastic processes), or modify it (inelastic processes). For a BEC in state $\ket{-}$, inelastic collisions are energetically forbidden and only elastic processes remain. As shown in Fig. \ref{fig1}(b), they can be described through an effective scattering length $a_{\text{-\,-}}$ which depends on the scattering properties of the bare states and on the composition of the system. In contrast, for a BEC in state $\ket{+}$ both elastic and inelastic processes are relevant.

\begin{figure*}[t]
\centering
\includegraphics[clip,scale=0.82]{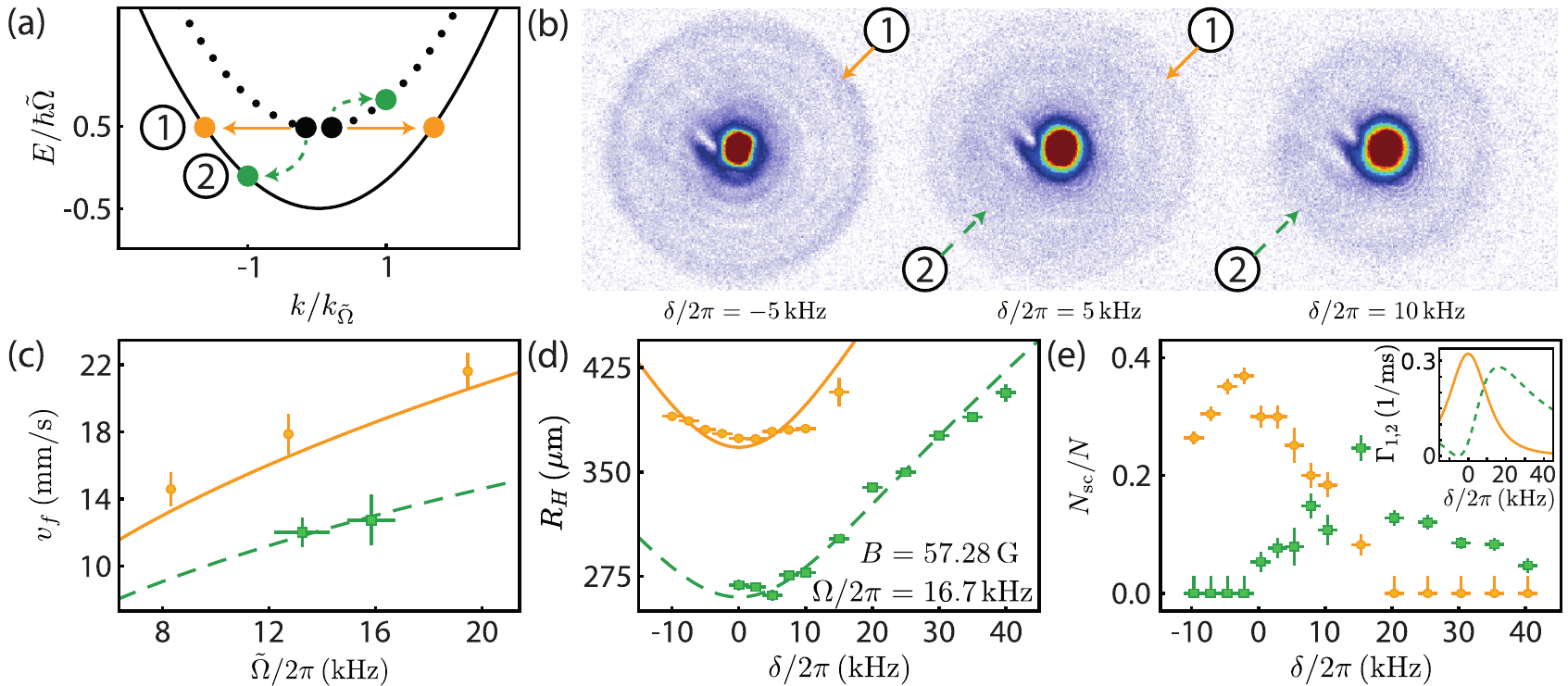}
\caption{Inelastic decay of the higher dressed state. (a) Sketch of possible dressed state changing collisions \circled{1}: $\ket{++}\rightarrow\ket{--}$ (orange, solid arrows) and \circled{2}: $\ket{++}\rightarrow\left(\ket{+-}+\ket{-+}\right)/\sqrt{2}$ (green, dashed arrows). Energy $E$ and momentum $k$ are expressed in terms of $\tilde{\Omega}=\sqrt{\Omega^2+\delta^2}$ and $k_{\tilde{\Omega}}=\sqrt{m\tilde{\Omega}/\hbar}$, respectively. (b) Measured momentum distribution of the collision products. Images correspond to the average of $10$ independent measurements. The likelihood of processes \circled{1} and \circled{2} depends on $\delta$. (c) Velocity of the scattered atoms $v_f$ vs. $\tilde{\Omega}$. (d) Radius of the halos $R_H$ for an expansion time $t_{\mathrm{exp}}=20.1$ ms. (e) Fraction of scattered atoms $N_{\mathrm{sc}}/N$ vs. $\delta$ \cite{NoteHalo}. Inset: Inelastic scattering rate $\Gamma_{1,2}$ vs. $\delta$ for $n = 1.3 \times 10^{14}$ atoms/cm$^3$. In (c), (d) and (e) orange circles (green squares) correspond to process \circled{1} (\circled{2}). Lines: theory predictions. Error bars: fit error (vertical) and uncertainty of $\delta$ and $\Omega$ (horizontal).}\label{fig2}
\end{figure*}

We implement these concepts with a $^{39}$K BEC at magnetic fields $B\sim56-57$ G. As atomic states we exploit the $m_F = - 1$ and $0$ magnetic sublevels of the $F = 1$ hyperfine manifold $\ket{\uparrow}\equiv\ket{F,m_F}=\ket{1,-1}$ and $\ket{\downarrow}\equiv\ket{1,0}$, for which the intra-state scattering lenghts are repulsive ($a_{\uparrow\uparrow}, a_{\downarrow\downarrow}>0$), and the inter-state scattering length is attractive ($a_{\uparrow\downarrow}<0$) \cite{Cabrera2018, Cheiney2018}. We coherently-couple the two states with an rf field. For all experiments its Rabi frequency $\Omega/2\pi>8$ kHz. We prepare single dressed states through Landau-Zener sweeps, starting from state $\ket{\uparrow}$ (unless explicitly stated otherwise) and ramping the detuning $\delta$ at a rate $\leq\,1$ kHz/$\mu$s to its final value \cite{NoteLandauZener}.

In a first series of experiments we focus on the elastic scattering properties of the lower dressed state $\ket{-}$. They are characterized by the effective scattering length $a_{\text{-\,-}}=a_{\uparrow\uparrow}\cos^4{\theta}+a_{\downarrow\downarrow}\sin^4{\theta}+\frac{1}{2} a_{\uparrow\downarrow}\sin^2{2\theta}$, and thus depend on the state composition of the system \emph{via} $\delta$ \cite{Search2001, NoteScattering}.

We experimentally probe this dependency by performing expansion measurements in an optical waveguide. To this end, we prepare a BEC in state $\ket{-}$ with $\Omega/2\pi=20.0(6)$ kHz and variable detuning $\delta$ using a ramp rate of $0.83$ kHz/ms. The magnetic field is set to $B=57.280(2)$ G, for which $a_{\uparrow\uparrow}/a_0=32.5$, $a_{\downarrow\downarrow}/a_0=109$, $a_{\uparrow\downarrow}/a_0=-52.9$, and we always have $a_{\text{-\,-}}>0$. Here $a_0$ is the Bohr radius. After holding the gas for $5$ ms at the final detuning, we switch off the axial confinement abruptly, allowing it to expand for $21$ ms along a single-beam optical dipole trap (radial frequency $\omega_r/2\pi=133(1)$ Hz). We finally image the gas \emph{in situ} using a polarization phase contrast scheme \cite{Cabrera2018, Cheiney2018} and exploit the axial size of the cloud $\sigma_x$ after expansion to infer the scattering length $a_{\text{-\,-}}$ \cite{NoteThomasFermifit}. In the Thomas-Fermi regime the two are related by $a_{\text{-\,-}}\propto\sigma_x^5/N$ \cite{Castin1996}. Although this approximation is not strictly valid for all of our experimental parameters, we have verified by numerical solution of the time-dependent Gross-Pitaevskii equation (GPE) that estimating $a_{\text{-\,-}}$ through this scaling law results in errors below our experimental uncertainties.

Figure \ref{fig1}(c) shows our determination of $a_{\text{-\,-}}$ for various detunings (circles), corresponding to different values of the polarization parameter $P$ (squares). We determine the latter by Stern-Gerlach separation of the bare states during time-of-flight expansion, from which we extract $P=\left(N_{\uparrow}- N_{\downarrow}\right)/ \left(N_{\uparrow}+ N_{\downarrow}\right)$. In order to correct for systematic errors in the measurement and compare the results to the scattering length $a_{\text{-\,-}}$, we have scaled $\sigma_x^5/N$ to yield $a_{\uparrow\uparrow}$ at large positive $\delta$. Whereas for large positive (negative) values of $\delta$ the effective scattering length should approach $a_{\uparrow\uparrow}$ ($a_{\downarrow\downarrow}$), we expect a minimum at $\delta/2\pi=6.5$ kHz ($P=0.31$) due to the attractive character of the inter-state interactions $a_{\uparrow\downarrow}<0$. This is in good agreement with the experimental measurements. The data at large negative $\delta$ are in fair agreement with the limit $a_{\downarrow\downarrow}$. It is also in this regime that the scaling law yields the largest discrepancies with the GPE simulations (diamonds) \cite{NoteFig1}. In conclusion, this method provides tunability of $a_{\text{-\,-}}$ by more than $100\,a_0$ without introducing additional loss mechanisms.

Next, we consider the scattering properties of the higher dressed state $\ket{+}$. There, besides elastic collisions, two-body inelastic collisions leading to a change of the two-particle dressed state are also allowed. For our typical experimental parameters  they limit the lifetime of the BEC to $\sim1$ ms. Fig. \ref{fig2}(a) sketches the two possible inelastic processes: \circled{1} $\ket{++}\rightarrow\ket{--}$ and \circled{2} $\ket{++}\rightarrow\left(\ket{+-}+\ket{-+}\right)/\sqrt{2}$. Both lead to the creation of correlated atom pairs with opposite momenta. They are accompanied by an energy release of either $\hbar\tilde{\Omega}$ or $\hbar\tilde{\Omega}/2$ per atom, corresponding to the energy gap between the two-particle dressed states. Similar processes occur in Raman-coupled BECs \cite{Williams2012}.

To reveal these dressed state changing collisions, we prepare rapidly (ramp rate $500$ kHz/ms) a pure sample of $\ket{+}$ atoms. We then immediately switch off the trap and let the gas expand for a time $t_{\mathrm{exp}}$. During the first $1$ ms the rf field is kept on, allowing us to characterize the decay processes in the absence of the trap. As depicted in Fig. \ref{fig2}(b), the time-of-flight images reveal the presence of halos of atoms expanding away from the condensate. Since atoms in a BEC  scatter with extremely low relative momenta, the halo radius $R_H$ at time $t_{\mathrm{exp}}$ directly reflects the velocity of the collision products $v_f=R_H/t_{\mathrm{exp}}$. Processes \circled{1} and \circled{2} can be distinguished because the velocities are given by $v_1=\sqrt{2\hbar\tilde{\Omega}/m}$ and $v_2=\sqrt{\hbar\tilde{\Omega}/m}$, respectively. Here $m$ is the mass of $^{39}$K. Experimentally, we observe that the likelihood of the two processes depends on the dressed state composition, and thus on $\delta$.

Figures \ref{fig2}(c), (d) and (e) present a more systematic study of these inelastic processes as a function of the parameters of the coupling field. Figure \ref{fig2}(c) depicts the velocity of the atoms in each halo vs. $\tilde{\Omega}$, determined by measuring $R_H$ for different values of $t_{\mathrm{exp}}$. Figure \ref{fig2}(d) shows the measured halo radius as a function of $\delta$. In both figures, circles (squares) correspond to process \circled{1} (process \circled{2}). The measurements are in excellent agreement with the theoretical predictions without any fitting parameters (solid and dashed lines).

The scattering cross section of the two processes strongly depends on detuning. This can be clearly seen in Fig. \ref{fig2}(e), where we plot the fraction of atoms scattered in each halo $N_{\mathrm{sc}}/N$ as a function of $\delta$ extracted from the same set of images as Fig. \ref{fig2}(d). The rate equation describing the evolution of the density in the initial state reads $\dot{n}=-2(\Gamma_1+\Gamma_2)n = -2 g^{(2)} (\sigma_1 v_1+\sigma_2 v_2) (n^2/2)$. Here $\Gamma_{1,2}$ are the inelastic scattering rates for processes \circled{1} and \circled{2}, $g^{(2)}=1$ is the BEC two-body correlation function, $n^2/2$ is the density of atom pairs, and $\sigma_{1}=\pi\left[\left(a_{\uparrow\uparrow}+ a_{\downarrow\downarrow}-2 a_{\uparrow\downarrow}\right)\sin^2{2\theta}\right]^2/2$ and $\sigma_{2}=4\pi\left[\left(a_{\uparrow\uparrow}\sin^2{\theta}-a_{\downarrow\downarrow}\cos^2{\theta}+ a_{\uparrow\downarrow}\cos{2\theta}\right)\sin{2\theta}\right]^2$ are the corresponding scattering cross sections \cite{Search2001, NoteScattering}. Our measurements agree qualitatively with the expected $\Gamma_{1,2}$ line shapes, see inset. For a quantitative prediction, the simultaneous reduction of $n$ due to the $1$~ms expansion of the cloud (which depends on $\delta$ \emph{via} $a_{\text{++}}$) must be taken into account. For $\delta\sim0$, the expansion can be neglected and $\sigma_2\sim0$, greatly simplifying the dynamics. In this regime, integration of the rate equation yields $N/N_{\mathrm{sc}}\sim 0.28$ for an initial density $n\sim 1.3\times10^{14}$ atoms/cm$^3$, in good agreement with the experiment.

After demonstration of the different collisional couplings present in dressed BECs, we refocus on the lower dressed state $\ket{-}$ and exploit the broad tunability of its effective scattering length to explore attractively interacting systems. In optical waveguides, this situation enables the study of bright solitons: matter-wave packets that propagate without changing their shape because attractive non-linearities balance the effect of dispersion along the unconfined direction. In coherently-coupled systems, these solitons are formed by dressed atoms: we call them dressed-state bright solitons. Compared to conventional bright solitons, they are bound by an additional mean-field attractive non-linearity which scales with density as an effective three-body force \cite{Note3body}. They are only stable while the gas is effectively one dimensional, with an interaction energy that remains below $\hbar\omega_r$ \cite{Perez1998, Salasnich2002, Carr2002}.

\begin{figure}[t]
\centering
\includegraphics[clip,scale=0.86]{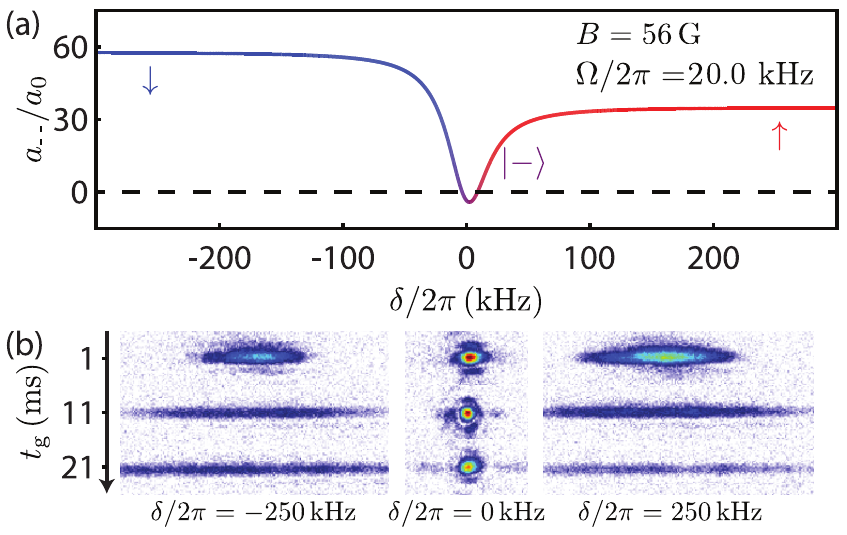}
\caption{Formation of a dressed-state bright soliton. (a) $a_{\text{-\,-}}$ vs. $\delta$. Near zero detuning $a_{\text{-\,-}}<0$. (b) \emph{In situ} dynamics of the gas after an evolution time $t_{\mathrm{g}}$ in the optical waveguide. For $\delta/2\pi=0$ ($a_{\text{-\,-}}/a_0=-3.5$) we observe the formation of a self-bound bright soliton, whereas for $\delta/2\pi=\pm250$ kHz interactions are repulsive and the gas expands.}\label{fig3}
\end{figure}

To observe this new type of bright soliton, we study the dynamics of a BEC in state $\ket{-}$ after release in an optical waveguide. The magnetic field is set to $B = 56.000(2)$ G, where $a_{\uparrow\uparrow}/a_0=35.1$, $a_{\downarrow\downarrow}/a_0=57.9$, $a_{\uparrow\downarrow}/a_0=-53.5$, and $a_{\text{-\,-}}$ can take negative values, see Fig. \ref{fig3}(a). We adiabatically prepare the system at different detunings (ramp rate $1$ kHz/ms). For $a_{\text{-\,-}}<0$ we keep the initial atom number below $N\sim3000$ to avoid collapse \cite{NoteSolitonPreparation}. We then remove the axial confinement in $15$ ms, allowing for free evolution in a waveguide. Fig. \ref{fig3}(b) shows \emph{in situ} images of the gas taken after an evolution time $t_{\text{g}}$. Whereas for $\delta/2\pi=\pm250$ kHz the gas expands, as expected for a repulsive BEC in states $\ket{\uparrow}$ or $\ket{\downarrow}$, for $\delta=0$ its shape remains unchanged. Here $a_{\text{-\,-}}/a_0=-3.5$ and we observe the formation of a single dressed-state bright soliton.

In the last series of experiments, we explore the response of the system to a quench of the effective scattering length from repulsive to attractive values. As demonstrated in recent experiments \cite{Nguyen2017, Everitt2017}, this triggers a modulational instability in the BEC: a mechanical instability where  fluctuations in the condensate density are exponentially enhanced by the attractive non-linearity. Consequently, the gas splits into several equally spaced components. The growth of the density modulation is dominated by the most unstable Bogoliubov modes, which have characteristic momentum $k_{\mathrm{MI}}\sim1/\xi$. Here $\xi=a_{\mathrm{ho}}/\sqrt{4 \left|a_{\text{-\,-}}\right|n_{1\mathrm{D}} }$ is the healing length of the BEC in the waveguide, $a_{\mathrm{ho}}=\sqrt{\hbar/m\omega_r}$ is the radial harmonic oscillator length, and $n_{1\mathrm{D}}$ is the line density of the system before the quench. The characteristic length and time scales of this process are $\lambda=2\pi/k_{\mathrm{MI}}$ and $\tau_{\mathrm{MI}}=2 m/\hbar k_{\mathrm{MI}}^2$, respectively. For $t>\tau_{\mathrm{MI}}$, each of the components evolves into a bright soliton, forming a soliton train \cite{Strecker2002, AlKhawaja2002, Salasnich2003, Carr2004a, Carr2004b}. For a system of size $L$ at the moment of the quench, the average number of solitons is expected to be $N_{S}= L/\lambda$ from simple length scale arguments \cite{Salasnich2003, Nguyen2017, Everitt2017}.

\begin{figure}[t]
\centering
\includegraphics[clip,scale=0.86]{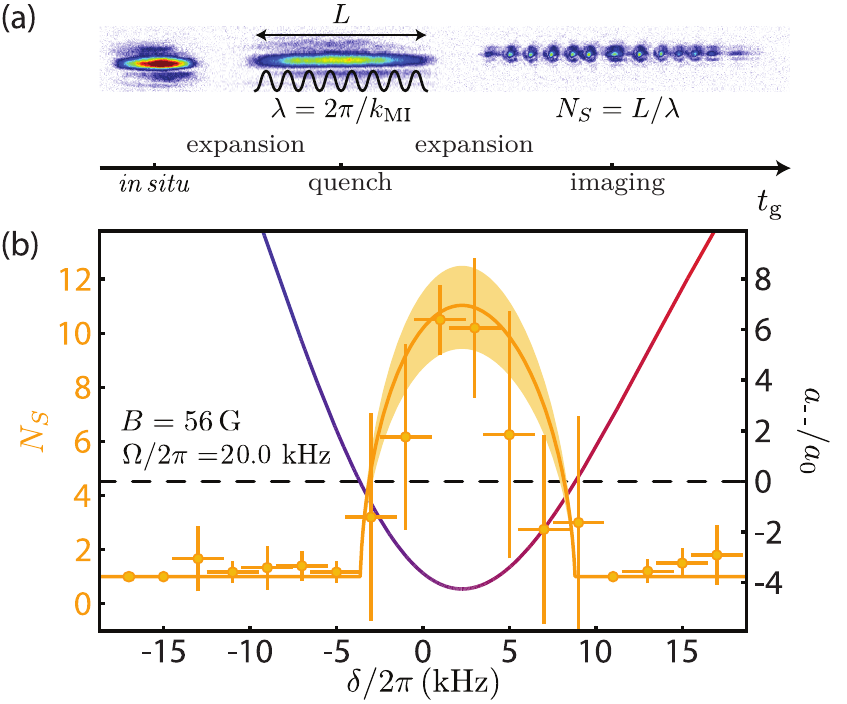}
\caption{Modulational instability and formation of bright soliton trains. (a) Sketch of the experimental sequence and exemplary \emph{in situ} images.
(b) Number of components observed per image $N_S$ vs. $\delta$ after the quench  (orange circles). Error bars: standard deviation of $4$ to $6$ independent measurements (vertical) and uncertainty of $\delta$ (horizontal). Orange line, left axis: theory prediction $N_S=L/\lambda$ (shaded area: uncertainty due to the systematic error in the atom number). Colored line, right axis: $a_{\text{-\,-}}$ (colorscale: value of $P$).}\label{fig4}
\end{figure}

Our experimental sequence is summarized in Fig. \ref{fig4}(a). The starting point of the experiment is a BEC of $65(15) \times 10^3$ atoms confined in a crossed optical dipole trap \cite{NoteTrapFrequencies}.
At $t=0$ we switch off the axial confinement and let the atoms expand in the waveguide (radial frequency $\omega_r/2\pi = 188(1)$ Hz) for $t_{\mathrm{g}}=11$ ms, reaching a size $L\sim 112\,\mu$m. At this point, we abruptly change $\delta$ (ramp rate $1$ kHz/$\mu$s), effectively quenching the scattering length from $35.1\,a_0$ to its final value. An additional expansion time of $10$ ms allows the development of the modulational instability and the formation of a soliton train \cite{NoteTimescales}, after which we image  the cloud \emph{in situ}.

Figure \ref{fig4}(b) shows the average number of components observed per image $N_S$ as a function of the final detuning \cite{NoteSolitonTrain}. Whereas the initial BEC has $N_S=1$, for all values of $\delta$ such that $a_{\text{-\,-}}<0$ we measure $N_S>1$. The maximum number of solitons in a train is observed at $\delta/2\pi=2.3$ kHz, which corresponds to the most attractive value of $a_{\text{-\,-}}$. Compared to previous experiments, where interactions were controlled using a magnetic Feshbach resonance \cite{Nguyen2017, Everitt2017}, our dressed-state approach enables ramp rates orders of magnitude faster \cite{NoteRampRates}.  This ensures a clear separation of timescales between the duration of the ramp and $\tau_{\mathrm{MI}}$, and allows us to perform experiments at more attractive interaction strengths. Our measurements show that the prediction $N_S=L/\lambda$ remains valid down to $a_{\text{-\,-}}/a_0=-4.2$ (solid line).

In conclusion, we have demonstrated fast temporal control of the collisional properties of rf-coupled $^{39}$K BECs. In the attractive regime, we have observed the formation of dressed-state bright solitons, and studied how the modulational instability triggered by an interaction quench develops into a bright soliton train. In future experiments, we could exploit dressed state changing collisions as a new source of correlated atom pairs \cite{Greiner2005, Spielman2006, Perrin2007, Buecker2011, Luecke2011}. We could also implement the coupling using optical Raman transitions, which would not only provide spatial control of the effective scattering length, but also allow engineering of higher partial wave collisions with tunable scattering amplitude \cite{Williams2012}. Concerning attractive non-linear systems, we could exploit the ability to perform fast interaction quenches to study soliton excitations and breathers \cite{DiCarli2019, Yurovsky2017}. We could also explore the three-body non-linearities predicted in rf-coupled BECs, which become important for smaller Rabi couplings. When stemming from quantum fluctuations, they are repulsive and should stabilize new types of quantum droplets \cite{Capellaro2017, Petrov2019}. Finally, due to their lack of Galilean invariance, Raman-coupled systems should enable the observation of chiral bright solitons \cite{Edmonds2013}.

We thank J. Brand, D. S. Petrov, A. Recati, and L. Santos for insightful discussions, A. Simoni for providing the scattering lengths of Ref. \cite{Roy2013} in numerical form, and A. Celi, E. Neri, and R. Ramos for a careful reading of the manuscript. We acknowledge funding from Fundaci\'{o} Privada Cellex, European Union (QUIC--641122), Ministerio de Ciencia, Innovaci\'{o}n y Universidades (QuDROP FIS2017-88334-P and Severo Ochoa SEV-2015-0522), Deutsche Forschungsgemeinschaft (Research Unit FOR2414, Project No. 277974659), and Generalitat de Catalunya (SGR1660 and CERCA program). J. S. acknowledges support from Ministerio de Ciencia, Innovaci\'{o}n y Universidades (BES-2015-072186), A. F. from La Caixa Foundation (ID 100010434, PhD fellowship LCF/BQ/DI18/11660040) and the European Union (Marie Sk{\l}odowska-Curie--713673), C. S. C. from the European Union (Marie Sk{\l}odowska-Curie--713729), C. R. C. from an ICFO-MPQ Cellex postdoctoral fellowship, and L. T. from Ministerio de Ciencia, Innovaci\'{o}n y Universidades (RYC-2015-17890).


\begin{thebibliography}{99}
\bibitem{NoteLandauZener} The detuning uncertainty is $\pm1.5$ kHz, limited by short-term magnetic field fluctuations ($\pm2$ mG). This restricts the ramp rates and Rabi frequencies accessible in our setup. The latter are calibrated independently through Rabi oscillations.
\bibitem{NoteThomasFermifit} The images are fitted with a 2D Thomas-Fermi profile to extract the axial size $\sigma_x$ and atom number $N$.
\bibitem{NoteHalo} We model the halo by a spherical shell with a Gaussian profile centered around the BEC and extract its parameters by fitting the time-of-flight images with its forward Abel transform.
\bibitem{Note3body} This term results from the density-dependent detuning appearing when $a_{\uparrow\uparrow}\neq a_{\downarrow\downarrow}$, which yields a differential mean-field frequency shift. For our experimental parameters, its effects are small and remain below our experimental resolution.
\bibitem{NoteTrapFrequencies} The initial trap frequencies are $(\omega_x,\omega_y,\omega_z)/2\pi=(26(1),190(1),188(1))$ Hz.
\bibitem{NoteTimescales} Empirically, this is a good compromise between $\tau_{\mathrm{MI}}\sim 5$ ms and the reduction of atom number at long times due to three-body recombination.
\end{thebibliography}
\end{document}